\documentclass[aps,pra,twocolumn,showpacs,amsmath,amssymb,floatfix,footinbib,superscriptaddress]{revtex4-1}

\usepackage{graphicx,natbib}
\usepackage{dcolumn}
\usepackage{bm}
\usepackage{amsmath,graphics,amssymb,psfrag}
\usepackage[dvipdfm,colorlinks=true,linkcolor=blue,urlcolor=blue,citecolor=blue]{hyperref}
\usepackage[left]{lineno}
\begin{document}

\title{\LARGE Monochromatic composite right/left handedness achieved in the quantized composite right/left handed transmission line }
\author{Shun-Cai Zhao}
\email[Corresponding author: ]{zhaosc@kmust.edu.cn }
\affiliation{Department of Physics, Faculty of Science, Kunming University of Science and Technology, Kunming, 650500, PR China}
\author{Xin Li }

\date{\today}

\begin{abstract}
The macro composite right/left handedness (CRLH) accompanies the positive/negative refraction index in the higher, microwave frequency bands in the composite right/left handed transmission line (CRLH-TL), respectively. In this paper, we adjust the refraction index of a quantized CRLH-TL via the squeezed parameters and the electronic components' parameters in the thermal squeezed state, and the refraction index shows the positive/negative jumping around the squeezed angle \(\varphi\)=\(\pi\) when it operates at a single frequency, and the similar result also arises while the refraction index is manipulated by the electronic components' parameters. The monochromatic CRLH achieved here breaks through the original definition in the macro CRLH-TL and provide a new implementation for the CRLH-TL.
\begin{description}
\item[PACS ]{78.20.Ci, 42.50.Gy, 81.05.Xj, 78.67.Pt }
\item[Keywords]{Monochromatic composite right/left handedness; quantized composite right/left handed transmission line; positive/negative refraction index}
\end{description}
\end{abstract}
\maketitle
\section{INTRODUCTION}
Since the first experimental demonstration of a left-handed (LH) structure with simultaneously negative effective permittivity
\(\varepsilon\) and permeability \(\mu\) \cite{1,2,3,4,5} in the microwave frequencies\cite{6}, some authors\cite{7,8,9,10}
adopted an engineering approach and developed the generalized transmission line (TL) theory to exhibit unprecedented features\cite{11,12,13,14,15}
in terms of performances or functionalities. These efforts resulted in the elaboration of the powerful CRLH concept, in which the right-handedness and left-handedness accompanying the positive/negative refraction index achived in a higher frequency band and the microwave frequency band, respectively. And the CRLH-TL led to a suite of novel guided-wave\cite{16}, radiated-wave\cite{17}, and refracted-wave devices and structures\cite{18,19}.

However, the technological advances of nanostructure manufacturing and the experimental achievements in manipulating quantum states have opened venues for novel devices and new findings in the miniaturization of circuits\cite{20,21,22,23,24}. In this context, a natural question is what's the role of quantum mechanical properties in the mesoscopic CRLH-TL when its compact size approaches the Fermi wavelength\cite{25,26}. Thus, the understanding of how the quantum mechanical behaviors alter the positive/negative refraction index\cite{27,28} in the quantized CRLH-TL\cite{29,30} is of crucial importance. Addressing these clearly requires a quantum description of the CRLH-TL, as is the main motivation of the present work.

Thus, we present a quantized description for the lossy CRLH-TL, and this quantized description is in line with the electromagnetic field quantization\cite{31} rather than the aforementioned work\cite{30}. And we focus on the refraction index of the CRLH-TL whose dissipation originates from the non-perfect conductivity of the conductors.
In this quantized CRLH-TL, the CRLH is implemented at a single frequency by the squeezed parameters and the electronic components¡¯ parameters in the thermal squeezed state. Here, the CRLH achieved in the quantized CRLH-TL is different from the macro CRLH-TL whose CRLH is obtained in a higher frequency band and the microwave frequency band, respectively. And the monochromatic CRLH may provide a new experimental direction for the CRLH-TL.

The overall layout of this paper is organized as follows. In Sec.2, we deduce the refraction index in the thermal squeezed state via the quantization of the traveling electric current wave in the unit cell circuit of the CRLH-TL, and we evaluate the refraction index dependent the squeezed parameters and the electronic components'
parameters in Sec.3. Sec.4 presents our conclusions and outlook.

\section{The refraction index of the quantized CRLH-TL in the thermal squeezed state }

\begin{figure}[!t]
\centerline{\includegraphics[width=0.45\columnwidth]{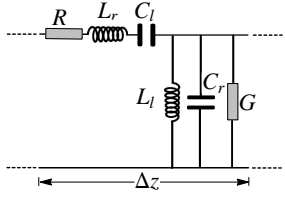}}
\caption{Schematic diagram of an unit-cell circuit for the dissipative quantized composite right-Left handed transmission line ( CRLH-TL) with the resistance \(R\) and conductance \( G\) representing the dissipation.}
\label{fig1}
\end{figure}
A segment of the dissipative CRLH-TL with the resistance \(R\) and conductance \( G\) representing the dissipation is shown in Fig. 1, and
the series capacitor \(C_l\) and inductance \(L_r\), shunt inductance \(L_l\) and capacitor \(C_r\) compose an equivalent unit-cell circuit. And the dimension \(\Delta z\) of the segment is much less than the wavelength at operating frequency. Starting from Kirchhoff's laws then to the telegraphist's equations, we obtain the electric current's dynamic differential equation (when \(\Delta z\rightarrow 0\)) as follows,
\begin{align*}\resizebox{0.65\hsize}{!}{\(
\frac{\partial^2{j(z)}}{\partial{z^2}}=[G+\frac{1}{i\omega{L_l}}+i\omega C_r]\times[R+\frac{1}{i\omega C_l}+i\omega L_r]j(z),\)} \tag{1}
\end{align*}

Where $j(z)$ is the current signal, and $\omega$ is the angle frequency. And the forward plane-wave solution to the above equation reads:
\begin{align*}
j(z)&=\exp{(-\sigma z)}(A e^{ i\beta{z} } + A^* e^{ -i\beta{z} }), \tag{2}
\end{align*}
in which \(\sigma\)=\(\sqrt{\frac{1}{2}(\sqrt{B^{2}+D^{2}}+B)}\),\(\beta\)=\(\sqrt{\frac{1}{2}(\sqrt{B^{2}+D^{2}}-B)}\).\\  They represent damping propagation coefficient and propagation constant, respectively. In the above expressions, \(B\)=\(G R-\omega^{2}C_{r}L_{r}+\frac{(L_r C_l+L_l C_r)\omega^{2}-1}{\omega^{2}L_l C_l}\), \(D\)=\(\omega (R C_{r}+L_{r}G) -\frac{G L_{l}+RC_{l}}{\omega C_{l} L_{l} } \). We consider the case of \(z_{0}\)=\(\frac{2\pi m}{\beta}\)\((m=0,1,2,3 \cdots)\), then the Hamiltonian of the equivalent unit-cell circuit can be written as via the integral in an unit-cell circuit,
\begin{align*}
H=\Gamma A^{*} A z_{0} \tag{3}
\end{align*}
with \(\Gamma\)=\(\frac{1+(R+L_{r})\omega^{2}C_{l}}{\omega^{2}C_{l}}\)+\(\frac{G\omega^{2} L_{l}^{2}+L_{l} }{(1+\omega L_{l}G+\omega^{2}L_{l}C_{r})^{2}}\)+\(C_{r}(\frac{1+\omega^{2} C_{l}L_{r}+R\omega C_{l}}{\omega C_{l}}\) \\+\(\frac{\omega L_{l}}{1+\omega L_{l}G+\omega^{2}L_{l}C_{r}})^{2}\). If \(A\) and \(A^{*}\) satisfy \(A =a\sqrt{\frac{\hbar\omega}{\Gamma z_{0}}} \) and \(A^{*}=a^{*}\sqrt{\frac{\hbar\omega}{\Gamma z_{0}}} \) with the commutation \([\hat{a},\hat{a}^{*}]=1\),
the quantum Hamiltonian of the unit-cell circuit can be written as \(\hat{H}=\hbar\omega \hat{a}^{\dag}\hat{a} \), which is almost identical with the quantum operator of the Hamiltonian of one-dimensional linear harmonic oscillator\cite{31}. And the Expression (2) can be quantized as,
\begin{align*}
\hat{j}(z)=\sqrt{\frac{\hbar\omega}{\Gamma z_{0}}}\exp{(-\sigma z)}(\hat{a} e^{i\beta{z}} + \hat{a}^{\dag} e^{-i\beta{z}})£¬\tag{4}
\end{align*}

The so-called thermal squeezed state  was defined as\cite{32,33,34}, \(|\alpha,z\rangle_{TSD}\)=\(\hat{T}(\theta)\hat{S}(z)\tilde{\hat{S}}(\tilde{z})\hat{D}(\alpha)\tilde{\hat{D}}(\tilde{\alpha})|0\tilde{0}\rangle\),
Generally, the operators \(\hat{F}(X)\), \(\tilde{\hat{F}}(\tilde{X})\) represent the operator in Hilbert space and its symmetric tilde space, respectively.
The creation operator and annihilation operator in the Hilbert space and its symmetric tilde space are in accordance with these commutation rules,
\([\tilde{\hat{a}}, \tilde{\hat{a}}^{\dag}]\)=1, and \([\tilde{\hat{a}}, \hat{a}]\)=\([\tilde{\hat{a}}, \hat{a}^{\dag}]\)=\([\hat{a}, \tilde{\hat{a}}^{\dag}]\)=\(0\).
\(\hat{T}(\theta)\)=\(\exp[-\theta(\hat{a}\tilde{\hat{a}}-\hat{a}^{\dag}\tilde{\hat{a}}^{\dag})]\), the thermal unitary operator is introduced by the thermal bogoliubov transformation\cite{33}. The displaced operator and the squeezed operator in the tilde space can be deduced according their corresponding operators \(\hat{D}(\alpha)\)=\(\exp(\alpha\hat{a}^{\dag}-\alpha^{*}\hat{a}) (\alpha\)=\(|\alpha|e^{i\delta},0\leq\delta\leq2\pi) \) , \(\hat{S}(z)\)=\(\exp[\frac{1}{2}(z^{*}\hat{a}^{2}-z\hat{a}^{\dag 2})]\)(z=\(\xi e^{i\varphi},0\leq\varphi\leq2\pi)\) in the Hilbert space.
Where $\alpha$, $\varphi$ and $\xi$ are a displaced parameter, squeezed angle and squeezed radius. \(\sinh(\theta)\)=\(\sqrt{n_{0}}\) with
\(n_{0}\)=\([\exp(\frac{\hbar\omega}{k_{B}T})-1]^{-1}\) being the numbers of  thermal phonon, which  has a direct relation with the temperature of classical thermodynamics and Boltzman constant.

The quantum fluctuation of the current operator in the thermal squeezed state can be calculated via the formula, \(e^{\lambda \hat{A}}\hat{B}e^{-\lambda \hat{A}}=\hat{B}+\lambda[\hat{A},\hat{B}]+\frac{\lambda^{2}}{2}[\hat{A},[\hat{A},\hat{B}]] \),
\begin{align*}\resizebox{0.65\hsize}{!}{\(
\langle(\Delta \hat{j})^{2}\rangle=\frac{\hbar\omega(2n_{0}+1)}{\Gamma z_{0}\exp(2\sigma z)}[ \cosh2\xi-\sinh2\xi\cos(2\beta z+\varphi)]\)},\tag{5}
\end{align*}
We notice that \( \beta z\) is infinitesimal when \(z\) is a dimensionless, then we can solve the propagation constant \(\beta\) through the Taylor expansion.
Utilizing the \( N=\frac{c_0 \beta}{\omega}\)\cite{35} ( \(c_0\) is the light speed in vacuum), the formula for the refraction index of the travelling current wave can be derived in the mesoscopic CRLH-TL as follow:
\begin{align*}
\resizebox{0.75\hsize}{!}{\(n=c_{0}\frac{\Gamma  z_{0}\exp(2\sigma z) \langle(\Delta \hat{j})^{2}\rangle-\hbar\omega(2n_{0}+1) (\cos\varphi\sinh2\xi+\cosh2\xi)}{2 \hbar\omega^{2}(2n_{0}+1) z \sin\varphi\sinh2\xi}\)}.  \tag{6}
\end{align*}
\section{Results and discussions}

The CRLH corresponds to the posive/negative refraction index, i.e., the right-handedness accompanies the positive refraction index in the higher frequency band while the left-handedness indicates the negative refraction index in microwave frequency band of the macro CRLH-TL. In the following, we analyze the refraction index via the Eq.(6) when the dimension of the unit-cell circuit in the CRLH-TL has minimized towards atomic-scale. However, the analytical expression Eq.(6) is rather cumbersome corresponding to electronic components' parameters and to the parameters of thermal squeezed state. For simplicity, we follow the numerical approach to get the intuitionistic conclusions. Several basic parameters should be selected before the analysis. We consider one unit length (\(z_{0}=1\mu m\)) and the mesoscpic CRLH-TL length \(z\)=\(10 z_{0} \), and set the quantum fluctuation of the current \(\langle(\Delta\hat{j})^{2}\rangle\)=\(10^{-23}\), the numbers of thermal phonon \(n_{0}\)=\(9\). The electronic components' parameters are listed in Table 1 whose magnitudes are referenced to Ref.\cite{36}.

\begin{table*}[t]
\footnotesize
\caption{ Electronic components' parameters in the equivalent unit-cell circuit.}
\tabcolsep 6pt
\begin{tabular*}{0.9\textwidth}{c|c|c|c|c|c|c|c|c|c}
\hline
Figs.  & \(C_l (pF)\) & \(L_l (pH)\)  &\(C_r (pF)\)  &\(L_r (pH)\) &\(\omega(GHz) \) &\(\xi(\mu m) \) &\(\varphi(\pi) \)  &\(R(\mu\Omega) \)  & \(G(\mu S) \) \\
\hline
Fig.2(a) & 9.8          & 0.968          &  4.35         & 0.998     &  --             & 2                & --               &0.8        & 0.25  \\
Fig.2(b) & 12.5         & 0.485          &  4.35         & 0.568     & 4.2             &  --              &  --              & 0.8       &  0.25 \\
Fig.3(a) & 9.8          & 0.968          &  4.35         & 0.998     & 4.5             &  --              & 0.25             &  --       & 0.25   \\
Fig.3(b) & 9.8          & 0.968          &  4.35         & 0.998     & --              & --               & 0.25             & 0.8       & 0.25  \\
\hline
\end{tabular*}
\end{table*}

\begin{figure}[htp]
\center
\includegraphics[width=0.485\columnwidth]{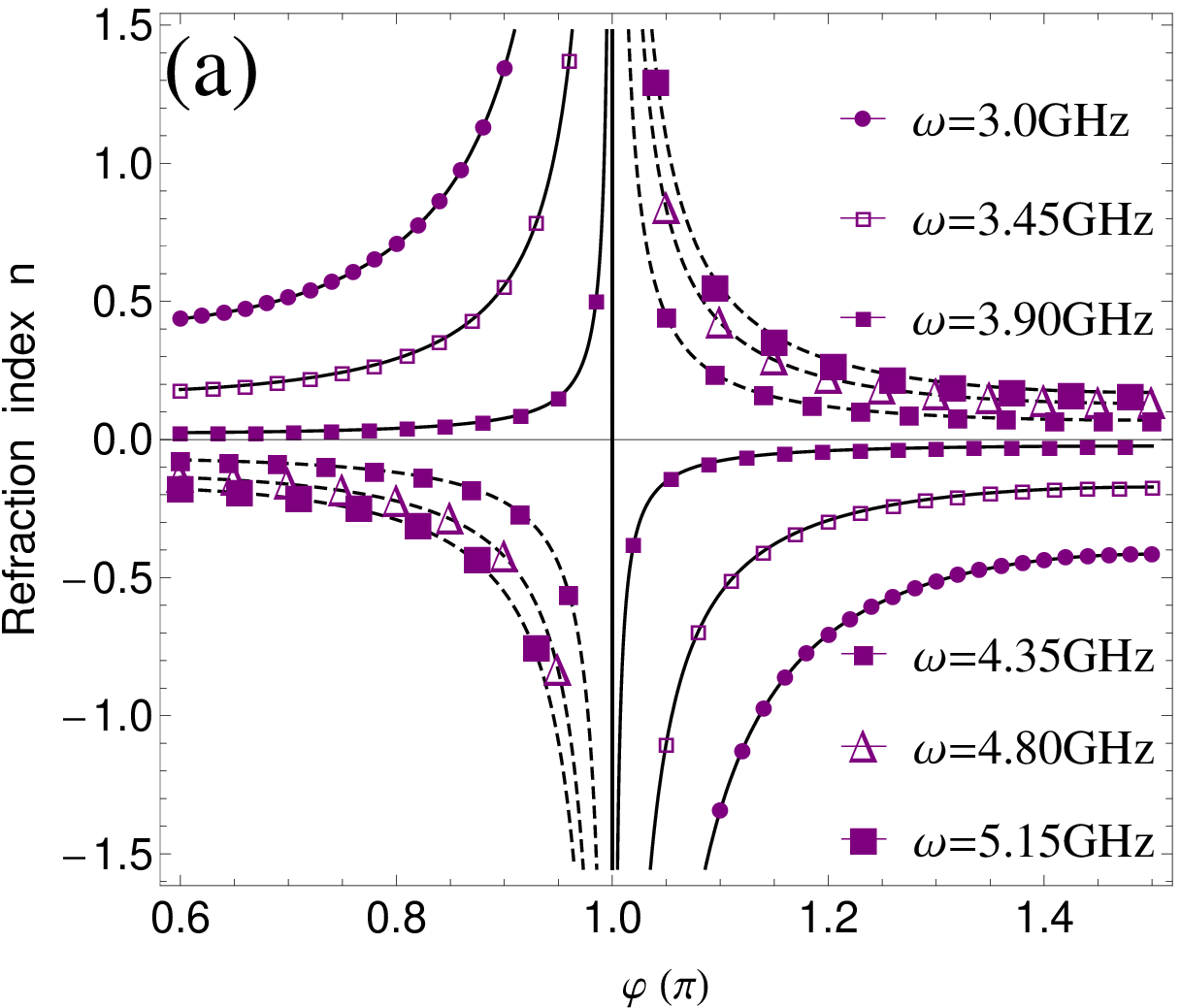 }
\includegraphics[width=0.45\columnwidth]{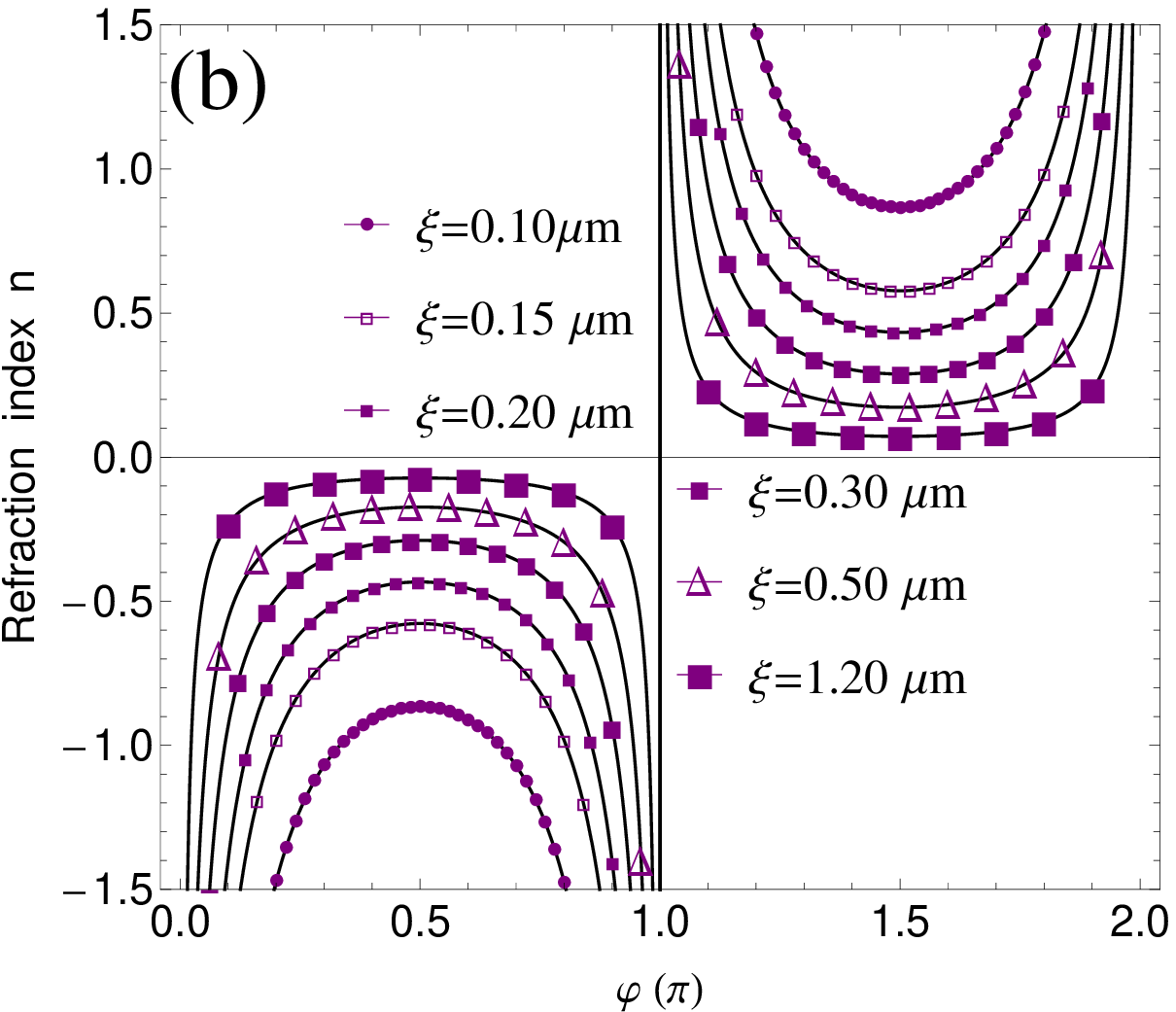 }
\caption{(Color online) The refraction index $ n $ versus the squeezed angle \(\varphi\) with different frequencies \(\omega\) in Fig.2(a), and with different squeezed radiuses \(\xi\) in Fig.2(b), respectively.}
\end{figure}\label{Fig.2}

The squeezed parameters involved the thermal squeezed state refer to the squeezed angle \(\varphi\) and the squeezed radius \(\xi\). Firstly, the refraction index $ n $ as the functions of the squeezed angle \(\varphi\) are plotted in Fig2.(a) and (b), respectively. It notes that refraction index $ n $ are negative in the range of [0.6\(\pi\), 1.0\(\pi\)] and positive in the range of [1.0\(\pi\), 1.5\(\pi\)] in Fig.2(a), respectively, and their corresponding angle frequencies are \(\omega\)=3.0GHz, 3.45GHz, 3.90GHz, respectively. However, the inverse case appears when the angle frequencies are \(\omega\)=4.35GHz, 4.80GHz, 5.15GHz in Fig.2(a). The negative refraction index appears in the range of [1.0\(\pi\), 1.5\(\pi\)] while the positive refractive index is in the range of [0.6\(\pi\), 1.0\(\pi\)]. And the squeezed angle \(\varphi\)=\(\pi\) is the demarcation point of the positive/negative refraction index both in the above two cases. In the macro CRLH-TL, the positive/negative refraction index appears in two different frequency bands, i.e., the high/microwave frequency bands, respectively. However, the jumping positive/negative refraction index operates at a single frequency \(\omega\) in this quantize CRLH-TL, which is the prominent difference between the quantize and macro CRLH-TL. The monochromatic CRLH achieved here breaks through the original definition in the macro CRLH-TL.

Fig.2(b) shows the refraction index \(n\) versus the squeezed angle \(\varphi\) with different squeezed radiuses \(\xi\), and the operating angle frequency in Fig.2(b) is set as \(\omega\)=4.2GHz. It shows the refraction index \(n\) are negative in the range of [0, 1.0\(\pi\)] and positive in the range of [1.0\(\pi\), 2.0\(\pi\)] when the squeezed radius \(\xi\) increases by the step of 0.05\(\mu\)m. The negative values of the refraction index \(n\) vary from 0 to -1.5. The curves show that \(\varphi\)=1\(\pi\) is the trigger point of the positive/negative refraction index. And the curves show the absolute values of the refractive index decrease when the squeezed radius \(\xi\) increases gradually. Fig.2(b) demonstrates the monochromatic CRLH can be flexibly adjusted by the squeezed radiuses \(\xi\).

\begin{figure}[htp]
\center
\includegraphics[width=0.45\columnwidth]{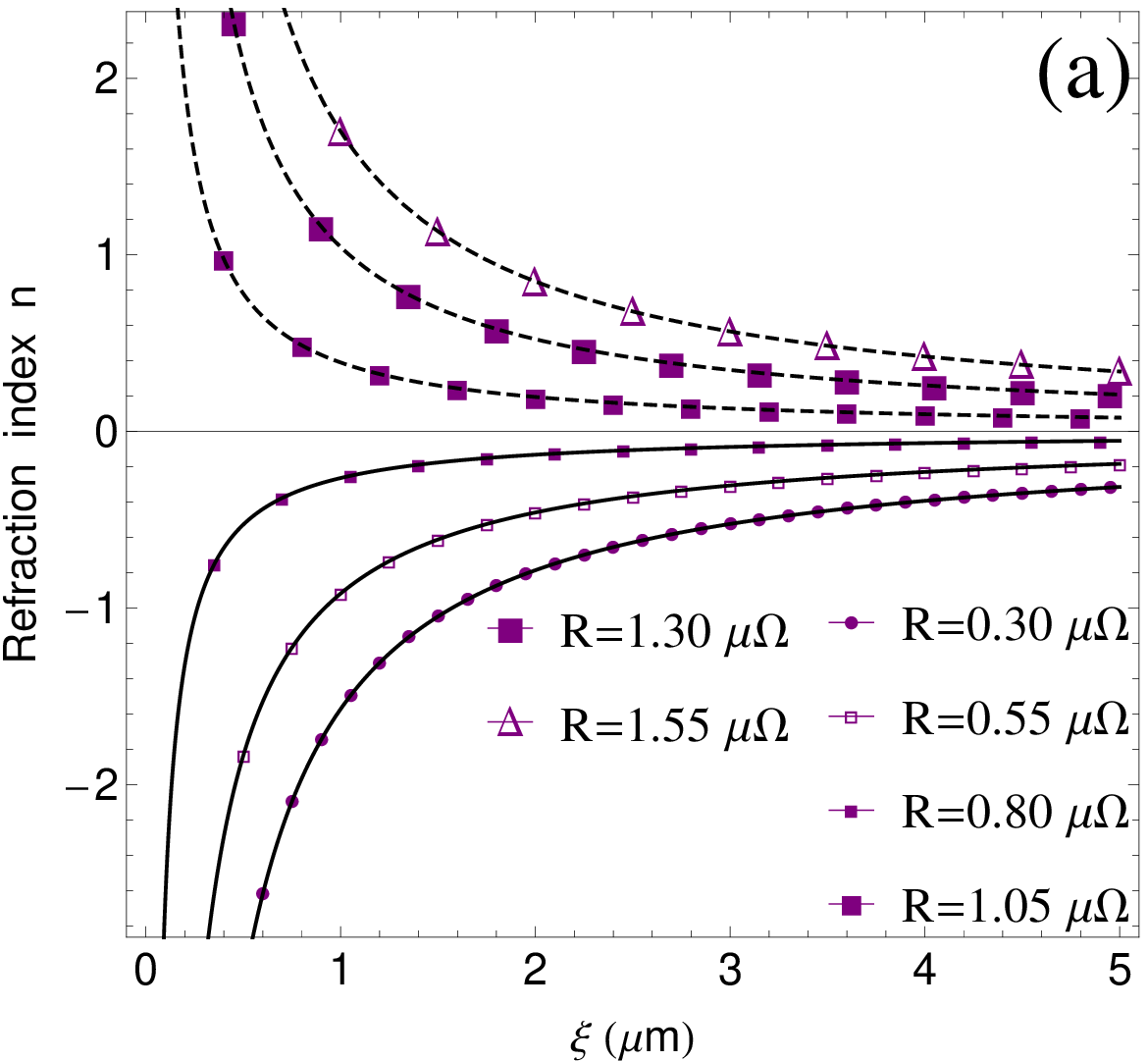 }
\includegraphics[width=0.47\columnwidth]{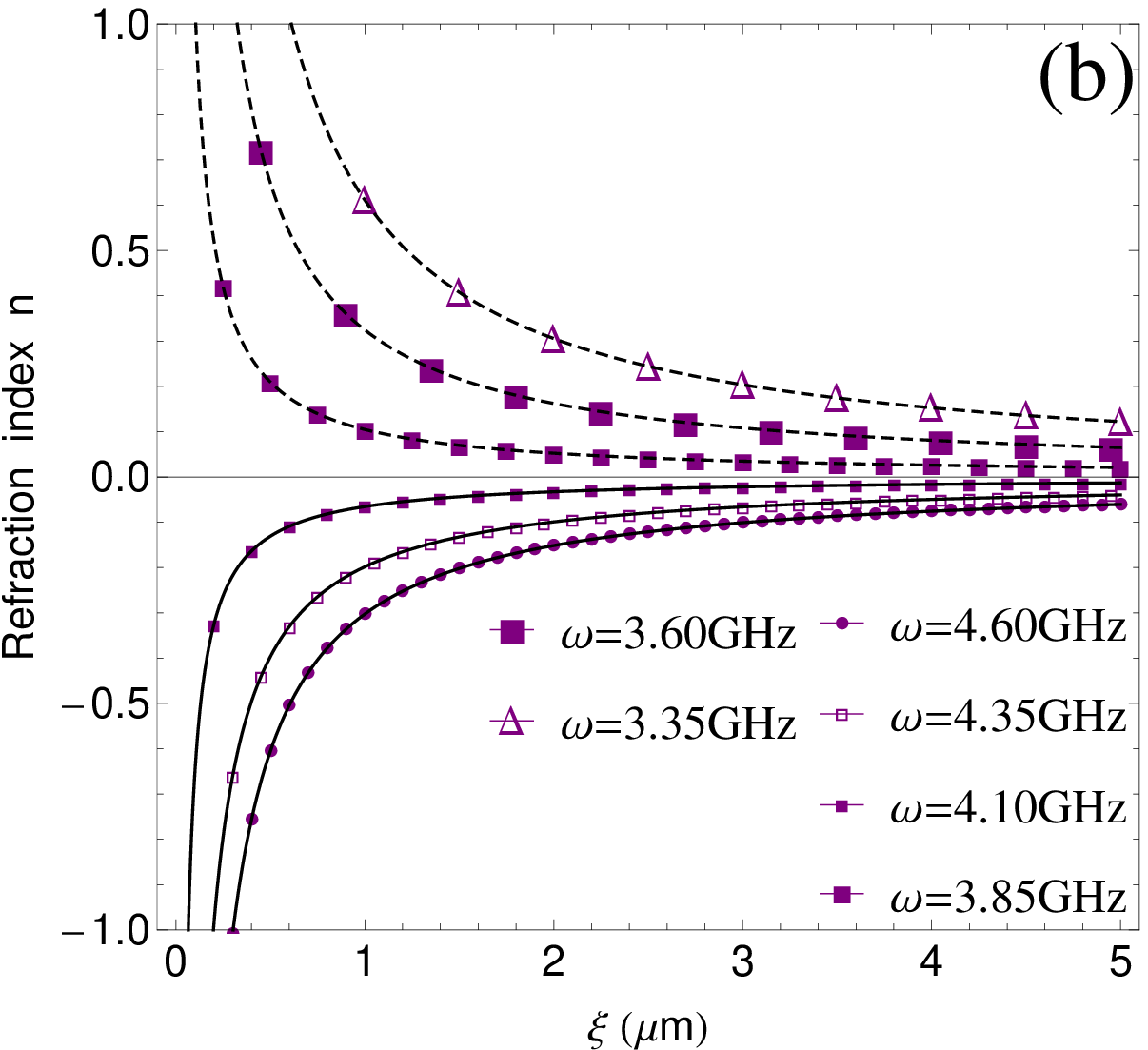  }
\caption{(Color online) The refraction index $ n $ versus the squeezed radius \(\xi\) with different resistances \(R\) in Fig.3(a), and with different angle frequencies \(\omega\) in Fig.3(b), respectively. }
\end{figure}\label{Fig.3}

However, the refraction index $ n $ versus the squeezed radius \(\xi\) shows different characteristics in Fig.3. In Fig.3(a) the angle frequency was set \(\omega\)=4.5GHz and the curves show the different features comparing to Fig.2. It shows the decreasing negative refraction index $ n $ in the increasing process of \(\xi\) when the value of resistance \(R\) \(\leq\)0.80 \(\mu\Omega\), and the decreasing positive refraction index $ n $ during the increasing \(\xi\) with  \(R\) \(>\)0.80 \(\mu\Omega\).
The result indicates that the smaller resistance \(R\) is facility to realize the negative refraction index. And the resistance \(R\)=0.80 \(\mu\Omega\) is the jumping point for the positive/negative refraction index, i.e. the monochromatic CRLH. In Fig.3(b), it shows the jumping positive/negative with different frequencies when the refraction index $ n $ is mediated by different values of the angle frequency \(\omega\). And we notice that the refraction index is negative with the higher angle frequency (\(\omega\)\(\geq\) 4.10GHz) while it gets the positive value with a lower angle frequency (\(\omega\)\(<\) 4.10GHz), which is similar to the macro CRLH-TL in different frequency bands. However, the CRLH corresponding to the positive/negative refraction index is in the lower, and higher frequency bands in this quantized CRLH-TL, respectively. Here, the multiple CRLH is shown in the quantized CRLH-TL.

\section{Conclusion and outlook}

In the present paper, we proposed a quantized CRLH in the thermal squeezed state and discussed its refraction index $ n $ via the squeezed parameters and the electronic components' parameters. When the refraction index is the function of the the squeezed angle \(\varphi\), the positive/negative refraction index jumping around the squeezed angle \(\varphi\)=\(\pi\) with single angle frequency, which indicates a monochromatic CRLH. While the monochromatic CRLH is also achieved when the refraction index is the function of the squeezed radius \(\xi\). And the positive/negative refraction index is be valued by whether the resistances \(R\)\(>\)0.80 \(\mu\Omega\) or not. However, the similar CRLH can also be obtained and the positive/negative achieved at a higher(\(\omega\)\(\geq\) 4.10GHz) and at the lower angle frequency(\(\omega\)\(<\) 4.10GHz), respectively. The quantized CRLH-TL breaks through the original definition of CRLH-TL and provides the multiple meanings of CRLH except for the monochromatic CRLH. These results may present a new direction for implementing CRLH in the context of CRLH-TL.
\begin{acknowledgments}
This work is supported by the National Natural Science Foundation of China (Grant Nos. 61205205 and 6156508508),
the General Program of Yunnan Applied Basic Research Project, China
( Grant No. 2016FB009 ) and the Foundation for Personnel training projects of Yunnan Province, China (Grant No. KKSY201207068).
\end{acknowledgments}
\bibliographystyle{alpha}

\end{document}